\documentclass{article}
\usepackage{iclr2027_conference}

\usepackage{hyperref}
\usepackage{url}
\usepackage{xcolor}
\usepackage{CJKutf8}
\usepackage{graphicx}
\usepackage{booktabs}
\usepackage{longtable}
\usepackage{array}
\newcolumntype{L}[1]{>{\raggedright\arraybackslash}p{#1}}

\title{JustGoal: With Only a Goal,\\the AI Baseline Is Above the Expert Kernels}

\author{
Yaojian Chen$^{1*}$, Dazhi Peng$^{2*}$, Zhui Zhu$^{1}$, Leshan Li$^{1}$, Haojun Sun$^{2}$,\\
\bf Yiyun Zheng$^{2}$, Qingfeng He$^{1}$, Shangzhan Li$^{3}$, Xianglong Deng$^{4}$, Yifei Shen$^{2}$,\\
\bf Shengye Xiang$^{1}$, Zhen Li$^{2}$, Lin Gan$^{1}$, Yuxuan Li$^{2\dagger}$, Xu Han$^{1\dagger}$, Zhiyuan Liu$^{1\dagger}$\\[0.6ex]
{\normalfont $^{1}$Tsinghua University \qquad $^{2}$ModelBest Inc. \qquad $^{3}$Harbin Institute of Technology}\\
{\normalfont $^{4}$University of Chinese Academy of Sciences}
}

\iclrfinalcopy
\begin{document}

\maketitle
\lhead{}
\renewcommand{\headrulewidth}{0pt}

{\renewcommand{\thefootnote}{}%
\footnotetext{$^{*}$Equal contribution.}%
\footnotetext{$^{\dagger}$Corresponding authors: Yuxuan Li, Xu Han, Zhiyuan Liu.}}

\begin{abstract}
A small number of GPU kernels often determines the efficiency of large
language model training and inference. High-performance kernel development
still typically draws on expert knowledge, from direct expert implementation,
to collaboration with general-purpose models, to specialized agents that
encode expert-designed evaluation methods, domain knowledge, and search
workflows. These approaches differ mainly in how expert knowledge enters the
development process. We ask how much additional value that knowledge still
provides for kernel optimization as foundation models grow stronger. To study
this question, we construct a goal-only baseline and call its concrete
experimental configuration JustGoal: a general-purpose agent receives
only the optimization objective and constraints, with no kernel-specific
evaluation infrastructure, knowledge base, or search workflow, and builds its
own evaluation loop. On NVIDIA B200 GPUs, JustGoal outperforms open-source
libraries on 8 of the 10 operators entering the comparison, with
per-operator geometric mean speedups of 1.14--2.34$\times$, and surpasses the
fastest publicly released specialized-agent kernel on 5 of the 9 operators
that have one. This marks a capability threshold: without task-time injection
of kernel-specific knowledge or optimization workflows, current frontier
models can already surpass human-maintained implementations on a range of
operators. In a controlled same-model comparison with Fable 5.1 on single
shapes, the KDA-0.5 harness reaches 0.95$\times$ the performance of JustGoal,
and single-mechanism ablations find no consistent benefit from domain
knowledge or predefined workflows. These results show that the effectiveness
of specialized machinery cannot be presumed from its complexity or from
results on earlier models. It should be established by a clear gain over a
goal-only baseline under the same model and task scope, and reevaluated when
the foundation model changes.
\end{abstract}

\section{Introduction}
\label{sec:intro}

The computational efficiency of large language model training and inference
depends critically on a small set of GPU kernels. Developing high-performance
implementations of these kernels requires jointly optimizing tiling strategies,
memory hierarchy utilization, asynchronous execution pipelines, and
hardware-specific instructions. These design choices are tightly coupled, and
their optimal configuration changes across hardware generations. Consequently,
the development of high-performance kernels has long relied on leading experts
devoting weeks or even months to manual optimization.
The FlashAttention series exemplifies this recurring investment: achieving high
performance on successive GPU architectures has required substantial redesigns
of its algorithms and execution pipelines
\citep{dao2022flashattention,shah2024flashattention3,dao2026flashattention4}.

Large language models have recently entered kernel development. Experts may
collaborate directly with general-purpose models. They may also encode
evaluation methods, domain knowledge, and search workflows into specialized
kernel agents such as KDA, AVO, AKO, and CAKE. These agents have matched or
surpassed expert-written implementations on selected operators
\citep{kda2026,avo2026,ako2026,cake2026}. What these approaches share is their
reliance on expert knowledge. As foundation models improve, capabilities that
previously had to be supplied explicitly may become native to the model. This
raises two related questions: how much additional value does expert knowledge
continue to provide for kernel optimization as model capabilities increase,
and how far can a model optimize kernels when given only the task objective and
constraints, with as little predefined expert knowledge as possible?

To examine these questions, we construct a goal-only baseline and call its
concrete experimental configuration JustGoal. It provides no kernel-specific
evaluation infrastructure, domain knowledge, or search workflow. A
general-purpose agent receives only the task objective and constraints. It
builds its own evaluation loop and optimization process. We observe the role of
expert knowledge from two measurable perspectives. First, we compare kernels
generated by JustGoal with human-maintained open-source libraries and publicly
released specialized-agent kernels, asking what performance level a goal-only
configuration already reaches. Second, under the same model and task scope, we
compare JustGoal with a specialized harness that adds evaluation
infrastructure, domain knowledge, and search workflows, directly measuring the
marginal value of these explicit expert mechanisms. Our experiments cover 11
operators spanning attention, linear attention, sparse attention, and
mixture-of-experts (MoE). We independently verify every kernel in the
comparison.

From the first perspective, JustGoal is faster than human-maintained
open-source libraries on 8 of the 10 operators entering the comparison, and
faster than the best publicly released specialized-agent kernel on 5 of the 9
operators that have one. This marks a capability threshold: without task-time
injection of kernel-specific knowledge or optimization workflows, current
frontier models can construct evaluation feedback, identify performance
bottlenecks, and surpass human-maintained implementations on a range of
operators. From the second perspective, in the scope-matched Fable 5.1
comparison, the full KDA-0.5 harness produces no overall gain on single-shape
tasks, and single-mechanism ablations show no consistent gain from domain
knowledge or predefined workflows. When the goal expands from optimizing one
shape to covering every workload in an operator definition, the harness still
provides a gain. To measure the net gain that the full specialized harness adds
beyond the model's own capabilities, it must be compared with a goal-only
baseline under the same model and task scope. This net gain should be
reevaluated whenever the foundation model changes.

\section{Related Work}
\label{sec:related}

\subsection{High-Performance Kernel Design}
\label{sec:related-kernel}

High-performance kernels must be co-designed around a model architecture and
target hardware. On the one hand, even when the operator remains unchanged,
changes in hardware resources and execution mechanisms can make an existing
implementation inefficient. On the other hand, evolving model architectures
introduce new operators and computational patterns. These two lines of
evolution jointly shape kernel design and continually create new development
and optimization demands.

\paragraph{Redesign driven by hardware evolution.}
High-performance kernel design requires aligning an algorithm's computational
and data movement demands with the resources of the target hardware. The
FlashAttention series exemplifies this hardware--algorithm co-evolution:
FlashAttention uses tiling to reduce data movement between high-bandwidth
memory (HBM) and on-chip SRAM \citep{dao2022flashattention}; FlashAttention-3
reorganizes computation and data transfers around Hopper's asynchronous
execution capabilities \citep{shah2024flashattention3}; and FlashAttention-4
further adapts the algorithm and execution pipeline to changes in the balance
of hardware resources, addressing bottlenecks such as exponential evaluation
and shared-memory access \citep{dao2026flashattention4}. This progression
illustrates that, even when an operator's mathematical definition remains
unchanged, achieving high performance requires experts to reassess bottlenecks
and adapt implementations to the target hardware.

\paragraph{New demands driven by model evolution.}
Evolving model architectures not only introduce new operators but also change
computational dependencies, state updates, and data-access patterns, making
existing kernels difficult to reuse directly. For example, DeepSeek provides
FlashMLA for multi-head latent attention (MLA) \citep{deepseek2025flashmla}.
With the introduction of DeepSeek Sparse Attention (DSA), it also provides the
corresponding sparse attention kernels through FlashMLA and implements the
indexer's scoring kernels in DeepGEMM
\citep{deepseek2025v32,deepseek2025deepgemm}. As a linear-attention
architecture, Gated DeltaNet (GDN) combines gating with the delta rule, and
Kimi Delta Attention extends it with a finer-grained gating mechanism. Both create computational patterns with
recurrent state updates \citep{yang2025gdn,kimiteam2025linear} that require
specialized implementations distinct from standard attention. Operator
implementations further branch into separate kernel paths for prefill and
decode, forward and backward, and different precision and accumulation
strategies. Sequence length and batch size further change the optimal design.
Continued model evolution expands the kernel space along multiple dimensions,
requiring sustained expert design and optimization for high-performance kernel
development.

\subsection{LLM-Based Kernel Generation}
\label{sec:related-agent}

Existing LLM-based kernel generation methods generally build on a
general-purpose agent and add specialized mechanisms for different challenges
in kernel optimization: execution feedback and search organize iteration,
independent validation ensures reliable evaluation, and domain knowledge and
diagnostic tools help models locate performance bottlenecks.

\paragraph{Execution feedback and search.}
KernelBench evaluates models' ability to generate faster, semantically equivalent
kernels from PyTorch reference implementations and explores iterative refinement
using execution and profiling feedback \citep{ouyang2025kernelbench}.
Existing work uses different search mechanisms to organize this optimization:
METR uses parallel tree search \citep{metr2025kernel}, AI CUDA Engineer uses
evolutionary search \citep{sakana2025cudaengineer}, K-Search guides search with a
co-evolving world model \citep{ksearch2026}, and AdaExplore distills recurring
failures into reusable validity rules \citep{adaexplore2026}. AVO introduces
autonomous agents into the mutation step of evolutionary search, producing
attention kernels that outperform expert implementations on some evaluated
configurations \citep{avo2026}.

\paragraph{Evaluation reliability.}
Optimization can exploit flaws in evaluation protocols. KernelBench-Verified
reports that a GPT-5.5-generated ReLU kernel simply returns its input for known
test shapes, yielding an apparent $374\times$ speedup
\citep{kernelbenchverified2026}. To guard against such failures, kernel agents
commonly incorporate independent validation, including sanitizer and anti-cheating
checks in AKO4X \citep{ako4x2026} and independent acceptance checks by Codex in
KDA \citep{kdareport2026}.

\paragraph{Domain knowledge and development environments.}
Timing reveals that a kernel is slow, but not why. Another line of work therefore
equips agents with domain knowledge and diagnostic tools. CudaForge uses a Judge
agent to interpret NCU metrics and guide revisions by a Coder agent
\citep{cudaforge2025}; KDA identifies the interface between models and the
engineering loop as a primary bottleneck, motivating KernelWiki, built from
production-code pull requests, an NCU profiling skill, and an IKET
(in-kernel event tracing) profiling skill \citep{kda2026,kdareport2026}; AKO4X maintains DSL-specific skills and a library
of operator pitfalls \citep{ako4x2026}; and CAKE lets agents directly edit typed
IR, with the compiler providing localized diagnostics before compilation
\citep{cake2026}.

\paragraph{General-purpose agent foundations and specialized layers.}
The mechanisms above are typically layered on the domain-independent tool use,
context management, and long-running autonomous execution provided by
general-purpose agents such as Claude Code and Codex
\citep{anthropic2025claudecode,openai2025codex}. KDA and AKO use Claude Code,
while AVO uses an internal general-purpose agent, and add kernel-specific
mechanisms for evaluation, knowledge, and workflow
\citep{kda2026,ako2026,avo2026}.
Table~\ref{tab:mechanisms} summarizes these mechanisms for the systems compared
in Section~\ref{sec:results}.

These approaches share a common pattern: each agent limitation motivates an
additional mechanism. Yet these mechanisms were designed and validated with the
underlying models available at the time; as those models improve, the marginal
value of each mechanism must be measured again. JustGoal removes these
kernel-specific layers and instantiates the goal-only baseline used for that
measurement.

\begin{table}[t]
\caption{Kernel-specific mechanisms added to general-purpose agent foundations.
KDA and AKO entries are based on their repositories and technical reports
\citep{kda2026,kdareport2026,ako4all2026,ako4x2026}; AVO and CAKE entries are based
on their papers \citep{avo2026,cake2026}.}
\label{tab:mechanisms}
\begin{center}
\small
\begin{tabular}{@{}lL{3.2cm}L{4.0cm}L{4.2cm}@{}}
\toprule
System & Evaluation & Knowledge & Workflow \\
\midrule
KDA & KDA-0.1 uses the FlashInfer-Bench standard; KDA-0.5 not released & KernelWiki (production PR corpus); NCU profiling skill; IKET (in-kernel event tracing) profiling skill & humanize2-based workflow; three-stage research--iterate--tune prompts; independent Codex acceptance checks at each step \\
\addlinespace
AVO & Scoring function: correctness checks against a reference and throughput measurement & Knowledge base: CUDA/PTX documentation, Blackwell specifications, and FA4 source code & Single-lineage evolution; only non-regressing versions are committed; supervisor intervention upon stagnation \\
\addlinespace
AKO & Compilation, correctness (including sanitizer), and timing benchmarks; anti-cheating checks & DSL-specific skills (including NCU profiling) and an operator-pitfall library (AKO4X) & Profile--modify--benchmark--commit loop with stagnation rules; multi-round campaigns with main and sub-agents (AKO4X) \\
\addlinespace
CAKE & Pre-compilation validation of typed IR, cost-model ranking, localized diagnostics, and CUPTI timing & Recurring failures distilled into validation rules, IR primitives, and optimization tactics & Generate--filter--evaluate--route stages; single-shape evolution followed by library generalization \\
\midrule
JustGoal & --- & --- & --- \\
\bottomrule
\end{tabular}
\end{center}
\end{table}

\section{JustGoal: With Only a Goal}
\label{sec:method}

\subsection{JustGoal}
\label{sec:justgoal}

\begin{figure}[t]
\centering
\fbox{\begin{minipage}{0.94\linewidth}
\small
This is the target operator we want: \{operator config\}. First measure its device time
on a B200. Starting from scratch, and never looking at or copying the
reference implementation, write a kernel that is at least 1.2$\times$ faster in
device time.
\begin{enumerate}
\item Use an isolated, clean environment; do not use existing skills, memory,
or other local files.
\item Make sure the kernel is correct. Preferably, first write a slow naive
fp32 implementation and compare it with the reference; their difference gives
a lower bound on the tolerance, which you multiply by a factor and
then use to compare your implementation against the slow fp32 implementation.
\item Do not ask questions; run in \texttt{-{}-dangerously-skip-permissions} mode.
\item Use the B200 cluster for the GPU.
\end{enumerate}
\end{minipage}}
\caption{The complete task input for JustGoal (English translation; the
original Chinese prompt is given in Appendix~\ref{sec:prompt_zh}).}
\label{fig:prompt}
\end{figure}

The box in Figure~\ref{fig:prompt} contains the complete input received by the
agent. Here, ``goal-only baseline'' denotes a control principle: give a
general-purpose agent only the objective and constraints, without adding
kernel-specific machinery. JustGoal instantiates this principle as the
experimental configuration used in this paper. It retains only the
general-purpose Claude Code agent substrate and access to the target
hardware, while removing the entire kernel-specific layer listed in
Table~\ref{tab:mechanisms}. The target hardware in all experiments in this
paper is B200. Item~1 requires a clean environment precisely to ensure this.

The model builds its own evaluation setup. The agent writes the compilation,
correctness-checking, and timing code within its optimization loop: it first
measures the reference kernel's device time as the target to surpass, then
writes a naive fp32 implementation as its correctness oracle.

We explicitly note that the prompt contains one piece of expert guidance:
item~2 instructs the agent to set the tolerance by multiplying the reference
implementation's error relative to the naive fp32 implementation by a factor.
This is a common numerical-validation practice and does not involve
kernel-optimization knowledge.

\subsection{Evaluation Protocol}
\label{sec:protocol}

We separate kernel generation from final acceptance. The evaluation in this
section is conducted independently by the authors after generation and
determines whether code is included in the reported results.

\paragraph{Baseline: open-source library kernels.}
Candidates are limited to human-maintained library implementations: the
official FlashAttention-4 kernels and FlashInfer (including FlashKDA). For each
shape, we select the fastest candidate that passes correctness checks and is
included in our evaluation.

\paragraph{Baseline: kernels generated by kernel agents.}
Candidates include kernels submitted by kernel agents to the MLSys'26 contest
and publicly released kernels from KDA-0.5, AKO4X, CAKE, and AVO
\citep{kda2026,ako4x2026,cake2026,avo2026}. The contest kernels are
likewise produced by kernel agents, so they belong to this category rather
than to the open-source libraries.
For each shape, we likewise select the fastest candidate that passes
correctness checks and is included in our evaluation. No publicly released
specialized-agent kernel for FA4 backward is available for this comparison.
Per-shape sources and selection details are provided in
Appendix~\ref{sec:released_latency}.

\paragraph{Correctness verification.}
FA4 uses the official FlashAttention standard; every other operator uses the
FlashInfer-Bench standard. Every implementation in the comparison must pass
the corresponding standard.

\paragraph{Timing methodology.}
All kernels are timed by the authors with FlashInfer-Bench in a uniform
environment. Before each measurement, the L2 cache is flushed (cold L2).
CUPTI records the interval from the start of the first kernel launched by the
entry point to the end of the last, excluding JIT compilation, cache flushing,
and synchronization overhead.

\section{Results}
\label{sec:results}

\subsection{Experimental Configuration}
\label{sec:config}

The models are Anthropic's Opus 4.8, Opus 5, and Fable 5.1, with reasoning
effort \texttt{high} throughout; the agent environment is Claude Code 2.1.304.
Kernel development and performance measurements run on NVIDIA B200 GPUs with
driver 580.105.08 and CUDA 13.0. Baselines and
timing follow the protocol of Section~\ref{sec:protocol}; agent-reported
measurements are not used as results. The
operators, shapes, and comparison groups of each experiment are given at the
start of its subsection.

\subsection{Overview: Comparison with Open-Source Libraries and Kernel Agents}
\label{sec:released_kernels}

This section gives an overview of JustGoal's results on 11 operators and 42
shapes and answers two different questions: whether JustGoal's kernels are
faster than human-maintained open-source libraries, and whether they are
faster than the kernels of specialized kernel agents. Each shape is one
session, with Fable 5.1 as the primary model; a few sessions were completed by
other models after a model fallback. We report geometric mean speedups per
operator; per-shape latencies, the models actually used, and baseline sources
are in Appendix~\ref{sec:released_latency}.

\begin{figure}[t]
\centering
\includegraphics[width=\linewidth]{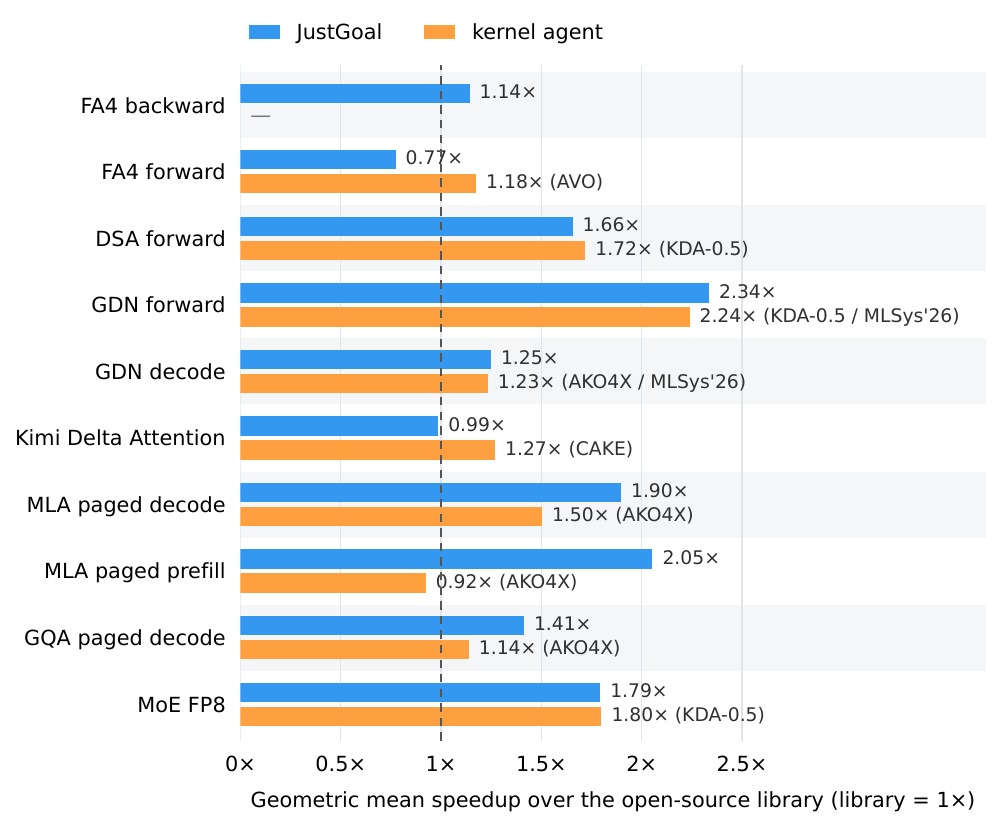}
\caption{Per-operator geometric mean speedup of JustGoal (blue) and kernel
agents (orange) over the open-source library; higher is better. The agent side
takes the fastest contest or released-agent kernel per shape; parentheses name
the sources. ``---'' indicates no comparable agent kernel. Per-shape details
are in Appendix~\ref{sec:released_latency}.}
\label{fig:released_kernels}
\end{figure}

\paragraph{Against open-source libraries.}
The blue bars in Figure~\ref{fig:released_kernels} are JustGoal's results
against open-source libraries. Excluding the DSA top-k indexer, whose baseline
protocol remains unverified, 10 operators enter the comparison, and JustGoal
is faster than the library on 8 of them, with
geometric means from 1.14$\times$ on FA4 backward to 2.34$\times$ on GDN
forward; it falls behind on FA4 forward (0.77$\times$) and Kimi Delta
Attention (0.99$\times$).

\paragraph{Against kernel agents.}
The orange bars are the kernel agents' kernels against the same open-source
library, taking for each shape the fastest of the MLSys'26 contest kernel and
the publicly released kernels of the four agent systems; the ratio of the blue
bar to the orange bar is therefore JustGoal's speedup over the agent kernel.
Nine of the 10 plotted operators have a comparable agent kernel (FA4 backward has
none); JustGoal is faster than the fastest agent kernel on five of them, ties
on MoE FP8 (1.79$\times$ versus 1.80$\times$), and falls behind on FA4 forward
(0.77$\times$ versus AVO's 1.18$\times$), DSA forward (1.66$\times$ versus
1.72$\times$), and Kimi Delta Attention
(0.99$\times$ versus CAKE's 1.27$\times$). These agent kernels were produced by
each system with its own choice of model and effort, so this comparison is
between the kernels themselves; what the harness itself contributes under the
same model is answered in Sections~\ref{sec:main_results} and~\ref{sec:scope}.

\paragraph{Relation to the released KDA-0.5 kernels.}
On the nine DSA, GDN, and MoE shapes, the kernels released in the KDA-0.5
repository reach a geometric mean of 1.29$\times$, and single-shape JustGoal
reaches 1.20$\times$ on the same nine shapes, close to the release; per-shape
values are in Appendix~\ref{sec:released_latency}. Our reproduction of its
harness with Fable 5.1 reaches 0.91$\times$ (same-GPU paired protocol,
Table~\ref{tab:main_models} in the appendix); the model and effort behind the
release are not public, and this gap bounds the fidelity of the reproduction.

The comparison with human-written libraries needs no further argument:
JustGoal is faster on 8 of the 10 operators, and faster than the fastest agent
release on 5 of the 9 operators that have one. Whether specialized machinery
adds value beyond this result is a separate question. The next section first
examines how the gap between full configurations changes across models; the
following scope-matched comparison directly measures the mechanism gain.

\subsection{Configuration Gaps Across Models}
\label{sec:main_results}

\paragraph{Configuration.}
Four operators (FA4 backward, DSA forward, GDN forward, MoE FP8), three shapes
each, 12 shapes in all; every run is repeated with each of the three models
under the same 24-hour budget. This section and Section~\ref{sec:scope}
compare four configurations:
\begin{itemize}
\item \emph{JustGoal, single shape}: one session optimizes one shape.
  12 shapes $\times$ 3 models, 36 sessions.
\item \emph{JustGoal, whole definition}: the same prompt, with the goal changed
  to every workload of the operator's FlashInfer-Bench definition (23 for DSA,
  100 for GDN, 19 for MoE FP8, 3 for FA4 backward); the session may dispatch by
  shape, and the final solution must pass the official evaluator on every
  workload. 4 operators $\times$ 3 models, 12 sessions.
\item \emph{KDA-0.5 reproduction}: the whole-definition scope, run through the
  KDA-0.5 harness with the same three models. KDA-0.5 has not fully released
  its prompts, so we follow its repository and technical report
  \citep{kda2026,kdareport2026}: the humanize2
  \texttt{flame\_chase} two-chaser workflow, each turn a fresh session sharing
  the git workspace; the KernelWiki, NCU report parsing, and
  \texttt{iket-profiling} skills; the public KDA 0.1 contest prompt as the
  task and FlashInfer-Bench as the evaluator. 4 operators $\times$ 3 models,
  12 runs.
\item \emph{KDA-0.5 reproduction, single shape}: the same KDA-0.5 harness;
  as in JustGoal, single shape, each run optimizes only one shape. Run with
  Fable 5.1 only: 12 shapes, 12 runs.
\end{itemize}
The reference kernel is the FA4 library kernel for FA4 backward and the
MLSys'26 contest kernel for the other three operators. Every implementation
follows the retest of the public repository. Two runs violated the task
constraints and are listed as is: the
Opus 4.8 JustGoal run on DSA heavy load read the reference implementation, and
the Opus 4.8 KDA-0.5 reproduction on MoE called FlashInfer library kernels.
The kernels released in the KDA-0.5 repository were produced
with an unknown model and are therefore not part of the same-model comparison;
their comparison with JustGoal is in Section~\ref{sec:released_kernels}.

\begin{figure}[t]
\centering
\includegraphics[width=\linewidth]{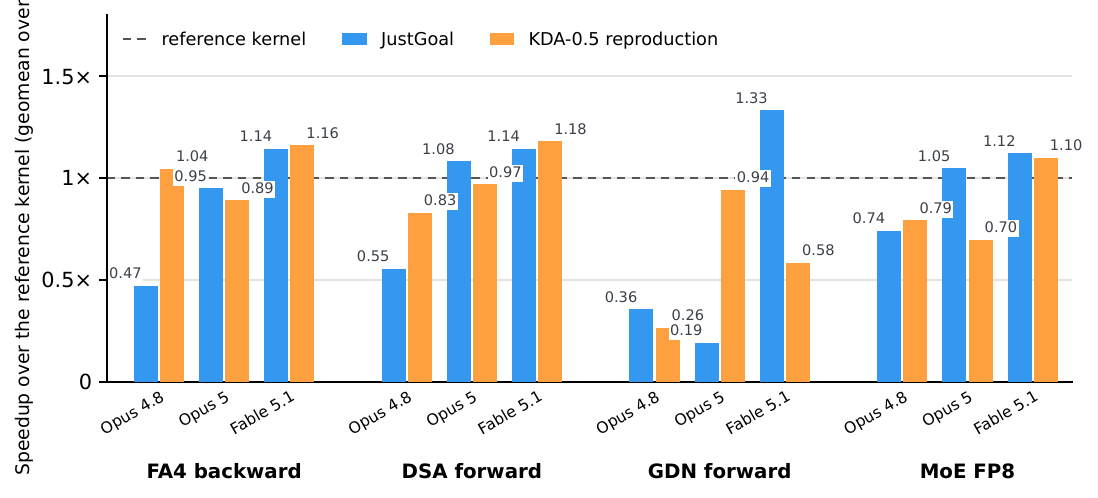}
\caption{Per-operator geometric mean speedup of single-shape JustGoal and the
KDA-0.5 reproduction (whole definition) over the reference kernel, by model;
the dashed line is the reference kernel.}
\label{fig:crossover}
\end{figure}

\paragraph{Results.}
The first three configurations exist for all three models.
Figure~\ref{fig:crossover} gives, per operator, the geometric mean speedup of
single-shape JustGoal and of the KDA-0.5 reproduction over the reference kernel
across models; per-shape values are in Table~\ref{tab:main_models} in the
appendix. The two differ in task scope, so this section only examines the
configuration gap; the scope-matched comparison is in Section~\ref{sec:scope}.

\paragraph{The gap between configurations changes across models.}
Over Table~\ref{tab:main_models}, single-shape JustGoal's geometric mean
speedup over the reference kernel rises from 0.51$\times$ with Opus 4.8 to
0.67$\times$ with Opus 5 and 1.18$\times$ with Fable 5.1 (12 shapes each), and
the same-model KDA-0.5 reproduction from 0.65$\times$ to 0.87$\times$ and
0.97$\times$: from Opus 4.8 to Fable 5.1, JustGoal improves by about
2.3$\times$ and the KDA-0.5 reproduction by about 1.5$\times$. JustGoal trails
the same-model KDA-0.5 reproduction with Opus 4.8 and Opus 5 and overtakes it
at Fable 5.1, where its geometric mean exceeds 1$\times$ on all four operators
while the KDA-0.5 reproduction stays below 1$\times$ on GDN. The two
configurations use different task scopes, so this comparison only shows that
their relative performance changes with the model; it cannot attribute the gap
to any mechanism in the harness.

\subsection{Harness and Task Scope: The Fable 5.1 2$\times$2}
\label{sec:scope}

To distinguish the influence of harness configuration from that of task scope,
we construct a 2$\times$2 comparison with Fable 5.1:
\{JustGoal, KDA-0.5 harness\} $\times$ \{single shape, full workload coverage\}.
Full workload coverage requires one run to cover every workload in the
operator's FlashInfer-Bench definition, allowing per-shape dispatch or
specialization. Table~\ref{tab:grid} reports per-operator geometric mean
speedups for the four configurations; per-shape results are in
Table~\ref{tab:main} in the appendix.

\begin{table}[t]
\centering
\caption{Harness and task scope with Fable 5.1: a 2$\times$2 comparison. Columns report per-operator geometric mean speedups over the reference kernels; the final row aggregates all 12 shapes.}
\label{tab:grid}
\small
\setlength{\tabcolsep}{4pt}
\begin{tabular*}{\linewidth}{@{\extracolsep{\fill}}lrrrr@{}}
\toprule
Operator & \multicolumn{2}{c}{single shape} & \multicolumn{2}{c}{full workload coverage} \\
\cmidrule(lr){2-3}\cmidrule(lr){4-5}
 & JustGoal & KDA-0.5 & JustGoal & KDA-0.5 \\
\midrule
FA4 backward & 1.14$\times$ & 1.28$\times$ & 0.79$\times$ & 1.18$\times$ \\
DSA forward & 1.14$\times$ & 0.94$\times$ & 0.74$\times$ & 1.20$\times$ \\
GDN forward & 1.33$\times$ & 1.13$\times$ & 0.69$\times$ & 0.57$\times$ \\
MoE FP8 & 1.12$\times$ & 1.17$\times$ & 0.64$\times$ & 1.11$\times$ \\
\midrule
Geomean & 1.18$\times$ & 1.12$\times$ & 0.71$\times$ & 0.97$\times$ \\
\bottomrule
\end{tabular*}
\end{table}

At single shape, JustGoal and the KDA-0.5 harness achieve geometric mean
speedups of 1.18$\times$ and 1.12$\times$ over the reference kernels,
respectively, so the specialized harness provides no overall improvement in
this setting. With full workload coverage, the corresponding speedups are
0.71$\times$ and 0.97$\times$, giving the KDA-0.5 harness a 1.37$\times$
advantage over JustGoal. The harness's relative advantage thus emerges as
task scope broadens, although both configurations achieve lower overall
performance than in their respective single-shape settings.

On the tested tasks, these results show that Fable 5.1 can already achieve
good performance when optimizing a single shape, but covering all shapes in
one run does not yet match the overall performance of independent per-shape
optimization. One possible explanation is that the model still struggles to
cluster shapes by their computational characteristics and decompose the
optimization task into subproblems that can be addressed separately.
This explanation requires further experimental validation: the present
comparison neither establishes the specific cause of the performance decline
nor attributes it to any individual harness mechanism. These observations are
limited to Fable 5.1; the other two models were not evaluated under all four
configurations, and their results are reported in
Table~\ref{tab:scope_models} in the appendix.

\subsection{Ablation Study}
\label{sec:ablation}

Fable 5.1 on three shapes, FA4 backward varlen, DSA forward heavy load, and GDN
forward long sequences, with default JustGoal as the reference and one
mechanism added at a time: a link to the KernelWiki repository in the prompt
(Kernel Wiki); an instruction to use NCU for performance analysis (NCU); a
report-parsing skill on top of the NCU instruction (NCU skill); and the goal
prefix replaced by the humanize2 \texttt{flame\_chase} two-agent workflow
(flame chase).

\begin{table}[htbp]
\centering
\caption{Speedup over the reference kernel under each configuration (higher is
better). The best per row is bold; $^{+}$/$^{-}$ marks a result faster/slower
than JustGoal.}
\label{tab:ablation}
\small
\setlength{\tabcolsep}{4pt}
\begin{tabular}{@{}lrrrrr@{}}
\toprule
Shape & JustGoal & Kernel Wiki & NCU & NCU skill & flame chase \\
\midrule
FA4 backward varlen & 1.08$\times$ & 0.91$\times$$^{-}$ & \textbf{1.28$\times$}$^{+}$ & 1.19$\times$$^{+}$ & 0.69$\times$$^{-}$ \\
DSA forward heavy & 0.90$\times$ & 1.12$\times$$^{+}$ & \textbf{1.17$\times$}$^{+}$ & 1.09$\times$$^{+}$ & 1.10$\times$$^{+}$ \\
GDN forward long & 1.32$\times$ & 1.14$\times$$^{-}$ & \textbf{1.65$\times$}$^{+}$ & 1.48$\times$$^{+}$ & 1.36$\times$$^{+}$ \\
\bottomrule
\end{tabular}
\end{table}

These ablations yield two observations: apart from NCU skill, the KDA-0.5
designs examined here show no consistent benefit, suggesting that the benefits
of specialized harnesses need to be reassessed with stronger models; NCU itself works well
without a dedicated skill, and adding one provides no further improvement.
This suggests that, with stronger models, the value lies in letting the model
know which tools are available; specialized mechanisms built around those
tools do not necessarily bring additional gains.

\paragraph{Accuracy verification.}
This comparison contrasts two kinds of accuracy constraint. Without a hard
constraint, JustGoal only asks for accuracy verification in the prompt and
says how to set the tolerance; the model writes the check itself, and nothing
outside it stops a kernel that misses the tolerance from becoming the final
version. With a hard constraint, as in the KDA-0.5 reproduction, the harness
runs the official evaluator in the loop and a failing kernel cannot be
submitted. Under the hard constraint the evaluator decides: all 12
reproductions with Opus 5 and Fable 5.1 pass the retest, and the 3 Opus 4.8
DSA kernels fail for not returning lse, a case the evaluator did not catch at
the time. Without it, Table~\ref{tab:passrate} counts, per JustGoal
configuration, the directories whose released version passes the retest of
the public repository: the model can build its own numerical check, yet some
kernels still fail accuracy. Single-shape JustGoal passes 11/12, 9/12 and
12/12 on the four main operators, 1/3, 4/7 and 22/26 on the other seven, and
the whole-definition runs 5/5, 4/5 and 5/5. The failures are accuracy,
missing outputs, build or runtime errors. With the
stronger model the model-built check approaches the pass rate of the hard
constraint but does not replace it.

\begin{table}[t]
\centering
\caption{Implementation directories whose released version passes the
independent retest / directories retested, per JustGoal configuration; all
generation-time correctness checks were built by the model.}
\label{tab:passrate}
\small
\begin{tabular}{@{}lrrr@{}}
\toprule
Configuration & Opus 4.8 & Opus 5 & Fable 5.1 \\
\midrule
JustGoal single shape, 4 main operators & 11/12 & 9/12 & 12/12 \\
JustGoal single shape, other 7 operators & 1/3 & 4/7 & 22/26 \\
JustGoal whole definition & 5/5 & 4/5 & 5/5 \\
\bottomrule
\end{tabular}
\end{table}

\subsection{Optimization Case Study}
\label{sec:case_study}

We use GDN (Gated DeltaNet) forward on long sequences (8192 tokens in 20
sequences of length 15 to 3371) to show how JustGoal crosses the contest kernel
from scratch. The reference is the MLSys'26 contest kernel from the KDA-0.5
submission, which the agent could only call as a black box; the audited
trajectory never reads its source. Fable 5.1 ran 8.9 hours without
intervention. Latencies are the session's own L2-flushed kernel span; the
reference is 125.1~$\mu$s. Appendix~\ref{sec:case_detail} gives the full record
and a second case, FA4 backward GQA.

\begin{figure}[t]
\centering
\includegraphics[width=\linewidth]{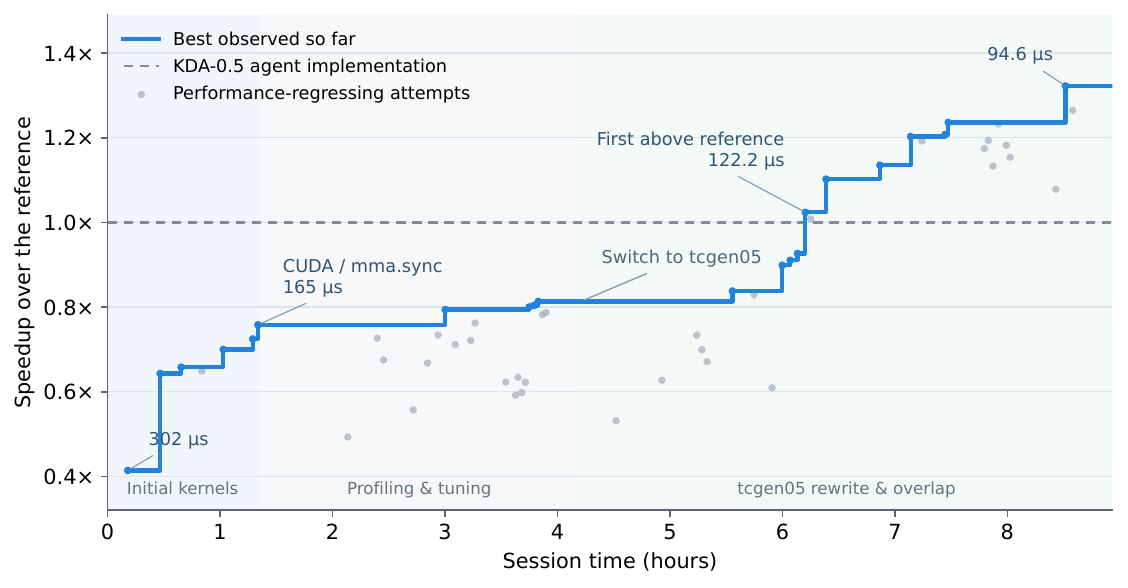}
\caption{Optimization trajectory of JustGoal (Fable 5.1) on GDN forward, long
sequences. The staircase is the session's best self-measured value; gray points
are log-reviewed change batches that did not improve it, and shading marks
strategy phases. Session measurements and the main table use different protocols.}
\label{fig:gdn_trajectory}
\end{figure}

The trajectory (Figure~\ref{fig:gdn_trajectory}) has three stages, each ending
when the agent measured that its toolchain had run out of headroom. Triton
started at 302~$\mu$s (0.41$\times$) and the agent put its floor near
146~$\mu$s. After microbenchmarking \texttt{mma.sync} on B200, it hand-wrote
CUDA kernels and, guided by NCU stall counters and in-kernel cycle counts,
reached 165~$\mu$s at 1.3 hours; further overlap and cluster work only reached
153.8~$\mu$s. At 4.2 hours it judged the path near the shared-memory bandwidth
limit and moved the recurrence to B200's tcgen05 tensor cores, checking
descriptor encodings and TMEM layouts with probe kernels. The first tcgen05
version was slower (235~$\mu$s); warp specialization, fusing the transposed O
product into the recurrence's MMA group, and running the intra-chunk kernel on
the remaining SMs crossed the reference at 6.2 hours. The final version
measures 95.6--97.3~$\mu$s over three processes (1.29--1.31$\times$;
1.32$\times$ under the paired re-measurement), with errors within 1.30 times
the reference's.

Profiling, toolchain switching and hardware probing were all the agent's own
choices. The prompt names no profiler, supplies no domain corpus and prescribes
no stages, whereas KDA-0.5 encodes the same behaviors as NCU and in-kernel
tracing skills, KernelWiki and three-stage prompts (Table~\ref{tab:mechanisms}). Given only
the goal, the model carried out the expert workflow on its own.

\section{Discussion}
\label{sec:discussion}

The value boundary of a specialized harness changes with foundation-model
capability and task scope. In the equal-scope Fable 5.1 comparison, the full
harness produces no overall gain on single shapes, and single-mechanism
ablations show no consistent benefit from domain knowledge or workflows. At
whole-definition scope, however, the harness still helps the model advance
more workloads simultaneously. This suggests that the primary value of
specialized machinery may be shifting from raising the peak performance of a
single target to helping models organize a broader task scope.
Task scope is also critical. The performance decline under full workload coverage shows that making one
implementation cover all workloads of an operator is a generality tax, and
current models cannot yet partition the workloads well enough on their own to
avoid it. This leaves two paths. One is to make the agent's own partitioning
ability a design goal of kernel agents and assess progress through the performance difference between single-shape
and full-workload configurations in Table~\ref{tab:grid}. The other is deep customization: in high-value scenarios
with fixed configurations, no single implementation has to cover every
workload, the generality tax disappears entirely, and the single-shape column
of Table~\ref{tab:main} is the level reached once it does. We call the latter
Forge Engineering. Each implementation optimizes performance only for its
target configuration but must remain correct for any input of that
configuration; the costs are optimization sessions that grow with the number
of configurations, plus per-configuration verification.

JustGoal surpasses expert implementations on some tasks studied in this paper,
suggesting that, as foundation models improve, the relative cost-effectiveness
of having experts hand-write every kernel from scratch is declining. Meanwhile,
the gain from a specialized harness varies with the model and task scope,
showing that expert mechanisms must be matched to the actual capabilities of
the foundation model. We believe a more promising path is to combine the depth
of expert knowledge with the model's broad prior knowledge, autonomous
experimentation, and scalable search to approach hardware performance limits.
Concretely, JustGoal can serve as a starting point that is updated with the
model: let the model explore independently first, then have experts analyze
optimization trajectories, profiling evidence, and failure cases to identify
specific limitations in bottleneck diagnosis, hardware knowledge, search
organization, or workload coverage, and add knowledge, diagnostic tools, or
constraints accordingly. The value of each added mechanism should still be
measured by its gain over JustGoal using the same model.

\section{Limitations}
\label{sec:limitations}

Our experiments focus on selected operators and shapes on NVIDIA B200 GPUs;
the applicability of our findings to other hardware and tasks remains to be
validated. Comparisons with publicly released kernels concern the kernels
themselves; the systems use different models and optimization budgets. The
same-model comparisons in Section~\ref{sec:main_results} separately examine the
harness. KDA has released its general workflow, base prompts, and related
skills, but the KDA-0.5 results repository primarily provides the final kernels
and retest code rather than the complete prompts, workflow configuration, and
search process used for that version. Our reproduction can therefore only
follow the available public materials and may differ from the original system.
KDA-0.5 also bundles domain priors with scope-organization mechanisms, and we
do not ablate these two groups separately. Their distinction in
Section~\ref{sec:scope} is our interpretation of the data.

Our correctness validation remains limited by test coverage. FA4 uses the
official FlashAttention standard, and every other kernel uses the
FlashInfer-Bench standard. Passing error-tolerance checks on a finite set of
test inputs does not guarantee correctness for all valid inputs and execution
conditions. The reported performance results should therefore be understood as
results obtained after passing the adopted validation procedures, rather than
a comprehensive guarantee of production readiness.

\section{Conclusion}
\label{sec:conclusion}

We first identify a change in current model capability. Given only objectives
and constraints, without predefined kernel-specific evaluation infrastructure,
domain knowledge, or search workflows, a general-purpose agent can construct
its own evaluation pipeline and iteratively optimize kernels. On NVIDIA B200
GPUs, JustGoal is faster than human-maintained open-source libraries on 8 of 10
operators and faster than the fastest specialized-agent release on 5 of 9.
Without task-time injection of kernel-specific knowledge or optimization
workflows, current frontier models can establish evaluation feedback, identify
performance bottlenecks, adapt optimization strategies, and cross the
performance line represented by human-maintained implementations on a range of
operators.

We further propose an evaluation principle that does not depend on the current
level of model capability. JustGoal is not a new agent architecture, but the
experimental configuration used to observe the performance level of a
goal-only setup and measure the marginal value of explicit expert mechanisms.
Specialized machinery should be considered effective only when it clearly
outperforms this baseline. When model capabilities change, the returns of
existing mechanisms should be measured again rather than inherited from
earlier results. The current model still benefits from a harness at broader
task scope; whether future models do so should also be answered by a
scope-matched comparison. We hope future work will not only apply new models to
existing systems but also use a goal-only configuration under the same model
and task scope as the baseline, reexamine each component's contribution against
it, and determine which components to retain, adapt, or remove.

\bibliographystyle{iclr2027_conference}
\bibliography{references}

\clearpage
\appendix
\section{Original Chinese Prompt}
\label{sec:prompt_zh}

Figure~\ref{fig:prompt} is an English translation; the box below is the original Chinese prompt given to the agent.

\begin{center}
\fbox{\begin{minipage}{0.94\linewidth}
\small
\begin{CJK}{UTF8}{gbsn}
这是我们要的目标算子（算子 config），请先在 B200 上测量它的 device time 时间。从 0 开始，绝对不看或抄写这个 ref 的代码实现，写出一个至少达到他 1.2X 倍快 device time 速度的 kernel 实现。
\begin{enumerate}
\item 请使用一个独立的干净的环境，不要用已有的 skills，memory，或者其它本机文件。
\item 请务必确保正确性。最好通过先写一个 fp32 的 naive 慢实现，再和 ref 比他们两的差这样可以获得一个下限容差范围再乘个值，然后用这个容差来对比你的实现和 fp32 慢实现的。
\item 不要问问题，用 \texttt{-{}-dangerously-skip-permissions} 模式。
\item GPU 请使用 B200 集群。
\end{enumerate}
\end{CJK}
\end{minipage}}
\end{center}

\section{Optimization Trajectories in Detail}
\label{sec:case_detail}

This appendix gives the full record behind Section~\ref{sec:case_study}: a
second case, FA4 backward GQA, and the step-by-step trajectory of the GDN
forward case summarized there. Both have complete trajectories, and all
per-version latencies are the sessions' own measurements.

\subsection{FA4 Backward GQA}
\label{sec:case_fa4}

We use FA4 backward GQA ($B=8$, $N=4096$, $H=16$, $H_{\mathrm{KV}}=2$,
$D=128$, causal, bf16). The reference is FlashAttention-4 by Tri Dao
et al.~\citep{dao2026flashattention4}, which the agent can invoke only as a
black box through its public API. We audited the complete trajectory and
confirmed that the agent did not read its source code. Fable 5.1 chose
Triton Gluon to implement the kernel and completed the task in four hours
without human intervention.

The agent first measured the reference kernel's device time and wrote a
naive fp32 implementation as a correctness oracle. Its first kernel took
2771~$\mu$s and passed correctness checks on the first attempt, but was
approximately 80\% slower than the reference. The agent then profiled stall reasons with NCU, added ablation
switches and \texttt{\%clock} instrumentation to measure cycles per stage.
It found that MMA alone required only about 830~$\mu$s, while softmax and
atomic accumulation of dQ accounted for most of the time. By reusing TMEM
and switching dQ accumulation to TMA bulk reduction, it reduced latency to
approximately 1320~$\mu$s after 2.5 hours.

\begin{figure}[t]
\centering
\includegraphics[width=\linewidth]{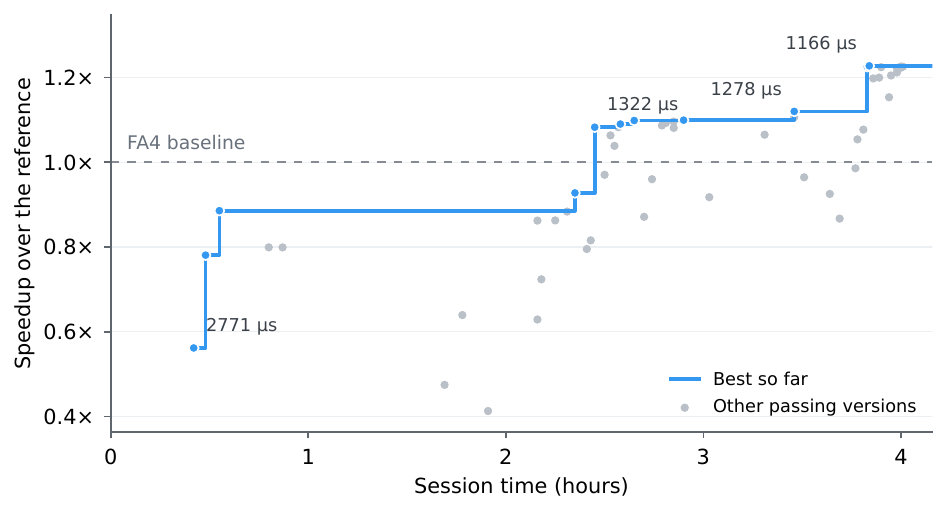}
\caption{Optimization trajectory of JustGoal (Fable 5.1) on FA4 backward GQA.
The figure uses session measurements; the final independent result is in the
main table.}
\label{fig:fa4_trajectory}
\end{figure}

Over the next 1.4 hours, more than a dozen new versions improved the best
latency only to 1278~$\mu$s (Figure~\ref{fig:fa4_trajectory}). The agent
wrote microbenchmarks and measured TMA reduction throughput of only
25--30 B/clk per SM, with roughly 2800 cycles required to release a staging
buffer. Disabling dQ reduction reduced the main kernel's latency to
931~$\mu$s. These observations pointed to the same cause: reduction reused
the staging buffer for dO, blocking dO loads. The agent therefore allocated
a dedicated 32 KB staging buffer for reduction and moved dO loads to drain
warps using \texttt{cp.async}, leaving TMA exclusively for reduction. This
change reduced latency to 1166~$\mu$s in a single step. Under the unified
cold-L2 paired re-measurement, the final kernel runs in 1321~$\mu$s against
1676~$\mu$s for FA4, a speedup of 1.27$\times$. The final kernel accumulates dQ in FP16 and rounds dS to BF16, so its internal numerics differ from FA4; its dQ, dK and dV all pass the gradient tolerance of the official FlashAttention standard, with errors at 20--33\% of the allowed budget.

Throughout this process, the agent independently used NCU, ablations, and
hardware microbenchmarks to diagnose bottlenecks and adjusted the kernel
structure accordingly, ultimately producing an operator that outperforms
the expert implementation.

\subsection{GDN Forward, Long Sequences}
\label{sec:case_gdn_detail}

The workload, reference and measurement protocol are those of
Section~\ref{sec:case_study}; this is the full record of the 8.9-hour run.

The agent first wrote a Triton version that passed correctness at once but
took 302~$\mu$s, 0.41$\times$ the reference (Figure~\ref{fig:gdn_trajectory}).
Sweeping tile and warp configurations showed a fixed per-step overhead in the
recurrence kernel and register spills in the intra-chunk kernel, with a floor
of about 146~$\mu$s for the Triton path, so it abandoned Triton, first wrote
microbenchmarks for \texttt{mma.sync} throughput and dependency latency on
B200, and then hand-wrote two CUDA C++ kernels with \texttt{mma.sync} and
\texttt{ldmatrix}, reaching 194.5~$\mu$s at 0.5 hours. It then checked spills
with \texttt{cuobjdump}, located hot spots with NCU stall and source counters,
and counted cycles per phase inside the kernel, bringing the recurrence from
4025 to 1993 cycles per step and the total to 165~$\mu$s at 1.3 hours. The
next two hours gained little: overlapping the three kernels on two streams
with per-block flags made things slower; several launch failures and
deadlocks were debugged with compute-sanitizer; thread-block clusters with TMA
multicast let 8 CTAs share one broadcast of the W/K tiles; and
\texttt{globaltimer} timelines of every CTA's work items showed a busy ratio
of only 0.74 with about 5~$\mu$s gaps between items, so it switched to static
round-robin assignment to remove the atomics. At 3.8 hours it stood at
153.8~$\mu$s, 0.81$\times$.

At 4.2 hours the agent reported plainly that correctness was met but
performance was not, gave a breakdown of where the time went, and judged that
the \texttt{mma.sync} path was near the shared-memory bandwidth limit, so the
recurrence had to move to B200's fifth-generation tensor cores (tcgen05, whose
operands are read from shared memory by the tensor core and whose accumulators
live in TMEM). It consulted the PTX ISA documentation and CUTLASS headers,
wrote probe kernels to verify descriptor encodings, swizzles and TMEM layouts,
and rewrote the recurrence kernel. The first tcgen05 version took
235~$\mu$s, about 3~$\mu$s per step, bottlenecked by 24 serial TMEM waits.
Over the next three hours it dumped TMEM intermediates and rebuilt O in fp32
to localize a numerical error, fused the transposed O product into the
recurrence's MMA group, microbenchmarked A operands from TMEM versus shared
memory, split TMA production, MMA issue and output write-back into dedicated
warps, and launched the recurrence first at the highest stream priority with
the intra-chunk kernel running on the remaining SMs. It first crossed the
reference at 6.2 hours (122~$\mu$s) and reached 101~$\mu$s at 7.5 hours.
Finally it made the intra-chunk kernel process one v head per CTA with the
longest sequences first, reaching 94.6~$\mu$s at 8.5 hours; the final version
measured 95.6 to 97.3~$\mu$s over three independent processes,
1.29--1.31$\times$ the reference, with output and state errors at 1.00 and
1.30 times the reference error. Under the same-GPU paired re-measurement
it reaches 1.32$\times$ over the MLSys'26 contest kernel shipped with
KDA-0.5.

\section{Per-Shape Latencies and Baseline Sources}
\label{sec:released_latency}

Table~\ref{tab:fable_shapes} lists implementation latencies and baseline sources.
The library column only lists human-maintained library implementations; the
MLSys'26 contest kernels are listed separately and, together with the released
agent kernels, form the kernel-agent baseline of
Section~\ref{sec:released_kernels}, taking the faster of the two per shape.
Some experiments were completed with other models due to model fallback:
the medium- and large-batch DSA top-k indexer tasks used Opus 5.5,
while the short and medium MLA paged prefill tasks and the medium- and
large-batch GQA paged decode tasks used Opus 5.
The target model for all other tasks was Fable 5.1.
The original length distributions of four shapes were not recovered, so they
were measured on the substitute distribution of an earlier round: GQA paged
decode medium and large batch, Kimi Delta Attention packed mixed lengths, and
MLA paged prefill large prefill.

The three DSA top-k indexer candidates passed correctness retests on their
respective limited inputs, with cold-L2 kernel spans of 47.59, 3531.5, and
13.25~$\mu$s; the large-batch result uses signed weights. The three historical top-k baselines (AKO4X
fast\_split\_v6 and v8, and Dogacel) compress scores to BF16 before selection
and fail a strict FP32 membership check on 9 of their 18 input
configurations. The agent kernel column gives the AKO4X result in the retest
repository, as a range over non-negative and signed weights; Dogacel measures 1.44--1.47,
15.24--24.46 and 18.02--27.43~$\mu$s. The library and contest columns remain
historical values. None of these baselines was paired with the candidates on
the same GPU, so the latencies are retained as audit records and are not used
to compute speedups in Figure~\ref{fig:released_kernels}.

Absolute latencies from different measurement batches are not comparable;
every speedup uses the baseline of its own pair. On GDN long sequences the
contest kernel measured 209.5~$\mu$s in the batch paired with JustGoal and
124.9~$\mu$s in the independent baseline retest; across the batches of this
appendix, the FA4 backward baseline spans 1371--1482, 1456--1676 and
1572--1706~$\mu$s on MHA, GQA and varlen MHA.

\begingroup
\scriptsize
\setlength{\tabcolsep}{3pt}
\begin{longtable}{@{}L{2.3cm}L{3.0cm}rL{2.2cm}L{2.1cm}L{2.1cm}@{}}
\caption{Per-shape latency ($\mu$s, lower is better) and baseline sources. The
library column contains only human-maintained implementations; contest and
released-agent kernels together form the kernel-agent baseline in the main text.}
\label{tab:fable_shapes}\\
\toprule
Operator & Shape & JustGoal & Library & Contest kernel & Released agent kernel \\
\midrule
\endfirsthead
\toprule
Operator & Shape & JustGoal & Library & Contest kernel & Released agent kernel \\
\midrule
\endhead
\midrule
\multicolumn{6}{r}{Continued on next page}\\
\endfoot
\bottomrule
\endlastfoot
FA4 backward & MHA & 1357.9 & 1481.9 (FA4) & — & — \\
 & GQA & 1321.2 & 1675.5 (FA4) & — & — \\
 & Variable-length MHA & 1581.9 & 1705.6 (FA4) & — & — \\
\addlinespace
FA4 forward & MHA (8,4096) non-causal & 874.0 & 981 (FA4) & — & 699 (AVO) \\
 & MHA (1,32768) non-causal & 6709.6 & 5930 (FA4) & — & 5270 (AVO) \\
 & MHA (8,4096) causal & 544.0 & 460 (FA4) & — & 395 (AVO) \\
 & MHA (1,32768) causal & 4188.9 & 3170 (FA4) & — & 2690 (AVO) \\
 & GQA g8 (8,4096) non-causal & 2277.2 & 1558 (FA4) & — & 1336 (AVO) \\
 & GQA g8 (1,32768) non-causal & 13205.8 & — & — & — \\
 & GQA g8 (8,4096) causal & 2501.4 & 960 (FA4) & — & 749 (AVO) \\
 & GQA g8 (1,32768) causal & 6667.6 & 5750 (FA4) & — & 5410 (AVO) \\
 & GQA g4 (8,4096) non-causal & 1767.0 & 1406 (FA4) & — & 1347 (AVO) \\
 & GQA g4 (1,32768) non-causal & 15886.9 & — & — & — \\
 & GQA g4 (8,4096) causal & 1003.0 & 929.5 (FA4) & — & 751 (AVO) \\
 & GQA g4 (1,32768) causal & 8321.2 & 6030 (FA4) & — & 5340 (AVO) \\
\addlinespace
DSA forward & Heavy workload & 11.42 & 13.71 (FlashInfer) & 10.27 (MLSys'26) & 10 (KDA-0.5) \\
 & Small request & 2.891 & 7.03 (FlashInfer) & 4.037 (MLSys'26) & 3 (KDA-0.5) \\
 & Mixed lengths & 7.115 & 11.07 (FlashInfer) & 8.427 (MLSys'26) & 7 (KDA-0.5) \\
\addlinespace
DSA top-k indexer & Single request, short sequence & 47.59 & 154.88 (DeepGEMM + FlashInfer) & 2.35 (MLSys'26) & 1.47--1.55 (AKO4X, FP32 fix) \\
 & Medium batch & 3531.5 & 315.59 (DeepGEMM + FlashInfer) & 17.21 (MLSys'26) & 17.39--17.70 (AKO4X, FP32 fix) \\
 & Large batch, mixed lengths & 13.25 & 313.28 (DeepGEMM + FlashInfer) & 21.73 (MLSys'26) & 24.24--24.49 (AKO4X, FP32 fix) \\
\addlinespace
GDN forward & Long sequences & 158.5 & 145.6 (FlashInfer) & 209.5 (MLSys'26) & 126 (KDA-0.5) \\
 & Single short sequence & 4.133 & 45.5 (FlashInfer) & 5.936 (MLSys'26) & 6 (KDA-0.5) \\
 & Few long sequences & 105.5 & 133 (FlashInfer) & 131.6 (MLSys'26) & 105 (KDA-0.5) \\
\addlinespace
GDN decode & Single request & 2.347 & 3.39 (FlashInfer) & 2.35 (MLSys'26) & 2.31 (AKO4X) \\
 & Medium batch & 4.725 & 5.56 (FlashInfer) & 5.27 (MLSys'26) & 4.75 (AKO4X) \\
 & Large batch & 11.91 & 13.66 (FlashInfer) & 12.51 (MLSys'26) & 12.79 (AKO4X) \\
\addlinespace
Kimi Delta Attention & Packed mixed lengths & 740.0 & 455.6 (FlashKDA) & — & 391 (CAKE) \\
 & Single fixed-length long sequence & 454.9 & 658.7 (FlashKDA) & — & 470 (CAKE) \\
 & Packed equal lengths & 506.7 & 545.1 (FlashKDA) & — & 435 (CAKE) \\
\addlinespace
MLA paged decode & Single request & 6.320 & 15.98 (FlashInfer) & — & 9.38 (AKO4X) \\
 & Medium batch, mixed lengths & 15.57 & 31.25 (FlashInfer) & — & 17.36 (AKO4X) \\
 & Large batch & 57.58 & 77.5 (FlashInfer) & — & 69.97 (AKO4X) \\
\addlinespace
MLA paged prefill & Short prefill & 4.981 & 22.886 (FA4) & — & 10.04 (AKO4X) \\
 & Medium prefill & 23.65 & 34.73 (FA4) & — & 22.29 (AKO4X) \\
 & Large prefill, mixed lengths & 430.6 & 553.0 (FA4) & — & 2491 (AKO4X) \\
\addlinespace
GQA paged decode & Single request & 4.853 & 7.488 (FA4) & — & 7.74 (AKO4X) \\
 & Medium batch, short sequences & 10.15 & 16.07 (FlashInfer) & — & 11.55 (AKO4X) \\
 & Large batch & 60.22 & 69.61 (FA4) & — & 63.47 (AKO4X) \\
\addlinespace
MoE FP8 & Single token & 50.09 & 86.41 (FlashInfer) & 57.15 (MLSys'26) & 49.05 (KDA-0.5) \\
 & Medium batch & 296.9 & 377.41 (FlashInfer) & 302.3 (MLSys'26) & 278.29 (KDA-0.5) \\
 & Large prefill & 967.1 & 2544 (FlashInfer) & 1173.6 (MLSys'26) & 1042.65 (KDA-0.5) \\
\end{longtable}
\endgroup

\subsection{Full Configuration Results for All Three Models}
\label{sec:main_models}

Table~\ref{tab:main} gives the per-shape speedups of the four Fable 5.1
configurations of Section~\ref{sec:scope}; Table~\ref{tab:grid} in the main
text is its per-operator geometric mean. Table~\ref{tab:main_models} gives the
per-shape speedups of three configurations under all three models; its
Fable 5.1 JustGoal column equals that of Table~\ref{tab:main}.
Table~\ref{tab:scope_models} gives the operator-level geometric means of the
three configurations under all three models; the two effects of
Table~\ref{tab:scope} are ratios of it.

\begin{table}[t]
\centering
\caption{Per-shape speedup of the four configurations over the reference
kernel with Fable 5.1 (higher is better). The fastest per row is bold;
per-operator geometric means are in Table~\ref{tab:grid}.}
\label{tab:main}
\small
\setlength{\tabcolsep}{5pt}
\resizebox{\linewidth}{!}{%
\begin{tabular}{@{}llrrrr@{}}
\toprule
Operator & Shape & JustGoal, single shape & KDA-0.5, single shape & JustGoal, whole definition & KDA-0.5, whole definition \\
\midrule
FA4 backward & MHA & 1.09$\times$ & \textbf{1.35$\times$} & 0.71$\times$ & 1.14$\times$ \\
 & GQA & 1.27$\times$ & \textbf{1.35$\times$} & 0.93$\times$ & 1.25$\times$ \\
 & varlen MHA & 1.08$\times$ & \textbf{1.16$\times$} & 0.75$\times$ & 1.14$\times$ \\
\addlinespace
DSA forward & heavy load & 0.90$\times$ & 1.05$\times$ & 0.76$\times$ & \textbf{1.36$\times$} \\
 & small request & \textbf{1.40$\times$} & 1.03$\times$ & 0.71$\times$ & 0.94$\times$ \\
 & mixed lengths & 1.18$\times$ & 0.78$\times$ & 0.74$\times$ & \textbf{1.35$\times$} \\
\addlinespace
GDN forward & long sequence & \textbf{1.32$\times$} & 0.92$\times$ & 0.70$\times$ & 0.53$\times$ \\
 & single short seq. & \textbf{1.44$\times$} & 1.41$\times$ & 0.90$\times$ & 1.03$\times$ \\
 & few long seqs. & \textbf{1.25$\times$} & 1.10$\times$ & 0.53$\times$ & 0.35$\times$ \\
\addlinespace
MoE FP8 & single token & 1.14$\times$ & \textbf{1.29$\times$} & 0.97$\times$ & 1.24$\times$ \\
 & medium batch & \textbf{1.02$\times$} & 0.90$\times$ & 0.64$\times$ & 0.98$\times$ \\
 & large prefill & 1.21$\times$ & \textbf{1.38$\times$} & 0.42$\times$ & 1.11$\times$ \\
\midrule
Geomean &  & \textbf{1.18$\times$} & 1.12$\times$ & 0.71$\times$ & 0.97$\times$ \\
\bottomrule
\end{tabular}}
\end{table}

\begin{table}[t]
\centering
\caption{Per-shape speedup in the same-model comparison (higher is better).
FA4 backward uses FA4 as the reference; the other operators use the MLSys'26
contest kernel. The faster configuration per model is bold. The scope-matched
comparison is in Table~\ref{tab:scope}.}
\label{tab:main_models}
\scriptsize
\setlength{\tabcolsep}{3pt}
\resizebox{\linewidth}{!}{%
\begin{tabular}{@{}lllrrrrrr@{}}
\toprule
Operator & Shape & Reference kernel & \multicolumn{2}{c}{Opus 4.8} & \multicolumn{2}{c}{Opus 5} & \multicolumn{2}{c}{Fable 5.1} \\
\cmidrule(lr){4-5}\cmidrule(lr){6-7}\cmidrule(lr){8-9}
 & & & JustGoal & KDA-0.5 & JustGoal & KDA-0.5 & JustGoal & KDA-0.5 \\
\midrule
FA4 backward & MHA & FA4 & 0.45$\times$ & \textbf{0.97$\times$} & \textbf{0.96$\times$} & 0.87$\times$ & \textbf{1.09$\times$} & 1.09$\times$ \\
 & GQA & FA4 & 0.42$\times$ & \textbf{1.17$\times$} & \textbf{1.12$\times$} & 0.92$\times$ & \textbf{1.27$\times$} & 1.26$\times$ \\
 & varlen MHA & FA4 & 0.55$\times$ & \textbf{1.01$\times$} & 0.79$\times$ & \textbf{0.88$\times$} & 1.08$\times$ & \textbf{1.14$\times$} \\
DSA forward & heavy load & contest kernel & 0.60$\times$ & \textbf{0.90$\times$} & 1.04$\times$ & \textbf{1.17$\times$} & 0.90$\times$ & \textbf{1.37$\times$} \\
 & small request & contest kernel & \textbf{0.96$\times$} & 0.68$\times$ & \textbf{1.07$\times$} & 0.73$\times$ & \textbf{1.40$\times$} & 0.91$\times$ \\
 & mixed lengths & contest kernel & 0.30$\times$ & \textbf{0.93$\times$} & \textbf{1.14$\times$} & 1.07$\times$ & 1.18$\times$ & \textbf{1.33$\times$} \\
GDN forward & long sequence & contest kernel & \textbf{0.58$\times$} & 0.26$\times$ & 0.14$\times$ & \textbf{1.02$\times$} & \textbf{1.32$\times$} & 0.53$\times$ \\
 & single short seq. & contest kernel & \textbf{0.38$\times$} & 0.30$\times$ & 1.02$\times$ & \textbf{1.05$\times$} & \textbf{1.44$\times$} & 1.09$\times$ \\
 & few long seqs. & contest kernel & 0.20$\times$ & \textbf{0.24$\times$} & 0.05$\times$ & \textbf{0.79$\times$} & \textbf{1.25$\times$} & 0.34$\times$ \\
MoE FP8 & single token & contest kernel & \textbf{1.09$\times$} & 0.96$\times$ & \textbf{1.18$\times$} & 0.83$\times$ & 1.14$\times$ & \textbf{1.18$\times$} \\
 & medium batch & contest kernel & 0.57$\times$ & \textbf{0.80$\times$} & \textbf{0.86$\times$} & 0.67$\times$ & \textbf{1.02$\times$} & 1.01$\times$ \\
 & large prefill & contest kernel & \textbf{0.65$\times$} & 0.65$\times$ & \textbf{1.12$\times$} & 0.60$\times$ & \textbf{1.21$\times$} & 1.12$\times$ \\
\midrule
\multicolumn{3}{@{}l}{single-shape JustGoal / KDA-0.5 repro., geomean} & \multicolumn{2}{c}{0.79$\times$ (12 shapes)} & \multicolumn{2}{c}{0.78$\times$ (12 shapes)} & \multicolumn{2}{c}{1.22$\times$ (12 shapes)} \\
\bottomrule
\end{tabular}}
\end{table}

\begin{table}[t]
\centering
\caption{Per-operator geometric mean speedup of the three configurations over
the reference kernel (higher is better). Single-shape values are computed from
Table~\ref{tab:main_models}; per-workload latencies are in
Appendix~\ref{sec:scope_us}.}
\label{tab:scope_models}
\small
\setlength{\tabcolsep}{4pt}
\resizebox{\linewidth}{!}{%
\begin{tabular}{@{}llrrr@{}}
\toprule
Operator & Model & JustGoal, single shape & JustGoal, whole definition & KDA-0.5 reproduction \\
\midrule
FA4 backward & Opus 4.8 & 0.47$\times$ & 0.50$\times$ & \textbf{1.07$\times$} \\
 & Opus 5 & 0.95$\times$ & 0.83$\times$ & \textbf{0.96$\times$} \\
 & Fable 5.1 & 1.14$\times$ & 0.79$\times$ & \textbf{1.18$\times$} \\
\addlinespace
DSA forward & Opus 4.8 & 0.55$\times$ & 0.45$\times$ & \textbf{0.67$\times$} \\
 & Opus 5 & 1.08$\times$ & 0.96$\times$ & \textbf{1.01$\times$} \\
 & Fable 5.1 & 1.14$\times$ & 0.74$\times$ & \textbf{1.20$\times$} \\
\addlinespace
GDN forward & Opus 4.8 & 0.36$\times$ & 0.13$\times$ & \textbf{0.18$\times$} \\
 & Opus 5 & 0.19$\times$ & 0.64$\times$ & \textbf{0.92$\times$} \\
 & Fable 5.1 & 1.33$\times$ & \textbf{0.69$\times$} & 0.57$\times$ \\
\addlinespace
MoE FP8 & Opus 4.8 & 0.74$\times$ & 0.32$\times$ & \textbf{0.63$\times$} \\
 & Opus 5 & 1.05$\times$ & 0.43$\times$ & \textbf{0.69$\times$} \\
 & Fable 5.1 & 1.12$\times$ & 0.64$\times$ & \textbf{1.11$\times$} \\
\bottomrule
\end{tabular}}
\end{table}

Table~\ref{tab:scope} reduces the results of the three models to three ratios,
each dividing the operator's geometric means over three shapes: scope effect
$=$ whole definition $\div$ single shape, below 1 is the loss from widening the
goal; harness effect $=$ KDA-0.5 reproduction $\div$ whole definition, above 1
means that the KDA-0.5 configuration is faster at equal scope; vs. single $=$ KDA-0.5
reproduction $\div$ single-shape JustGoal. In 11 of the 12 operator-model
combinations, whole-definition JustGoal trails the same-model KDA-0.5
reproduction, and it crosses the reference kernel in none. The harness and
scope effects are 1.72$\times$ and 0.61$\times$ with Opus 4.8, and
1.37$\times$ and 0.60$\times$ with Fable 5.1. These ratios describe gaps
between full configurations and do not isolate the contribution of any
individual mechanism.

\begin{table}[t]
\centering
\caption{Scope effect, harness effect, and gap to the single-shape ceiling per operator under the three models; last row: geometric mean.}
\label{tab:scope}
\small
\setlength{\tabcolsep}{4pt}
\resizebox{\linewidth}{!}{%
\begin{tabular}{@{}lrrrrrrrrr@{}}
\toprule
Operator & \multicolumn{3}{c}{Opus 4.8} & \multicolumn{3}{c}{Opus 5} & \multicolumn{3}{c}{Fable 5.1} \\
\cmidrule(lr){2-4}\cmidrule(lr){5-7}\cmidrule(lr){8-10}
 & scope & harness & vs. single & scope & harness & vs. single & scope & harness & vs. single \\
\midrule
FA4 backward & 1.07$\times$ & 2.13$\times$ & 2.27$\times$ & 0.87$\times$ & 1.16$\times$ & 1.01$\times$ & 0.69$\times$ & 1.49$\times$ & 1.03$\times$ \\
DSA forward & 0.81$\times$ & 1.50$\times$ & 1.21$\times$ & 0.89$\times$ & 1.05$\times$ & 0.93$\times$ & 0.65$\times$ & 1.63$\times$ & 1.05$\times$ \\
GDN forward & 0.37$\times$ & 1.41$\times$ & 0.52$\times$ & 3.32$\times$ & 1.45$\times$ & 4.81$\times$ & 0.52$\times$ & 0.83$\times$ & 0.43$\times$ \\
MoE FP8 & 0.44$\times$ & 1.97$\times$ & 0.86$\times$ & 0.41$\times$ & 1.59$\times$ & 0.66$\times$ & 0.57$\times$ & 1.73$\times$ & 0.99$\times$ \\
\midrule
Geomean & 0.61$\times$ & 1.72$\times$ & 1.05$\times$ & 1.02$\times$ & 1.30$\times$ & 1.32$\times$ & 0.60$\times$ & 1.37$\times$ & 0.82$\times$ \\
\bottomrule
\end{tabular}}
\end{table}

\subsection{Per-Workload Latencies of the Definition-Level Comparison}
\label{sec:scope_us}

Table~\ref{tab:scope_us} lists the per-workload latencies behind
Table~\ref{tab:scope}. All values were re-measured by the authors on one B200
with the official evaluator (\texttt{flashinfer-bench} main @
\texttt{40e6ca7}: CUPTI kernel span, cold L2, median, each definition's
official tolerances) in at least two alternating rounds; the whole-definition
solutions pass the official evaluator on every workload of their definitions,
and only the three shapes used in the main text are timed here. The KDA-0.5
reproduction is the version at the end of the run. Library column: FA4
4.0.0b19 for FA4 backward; for DSA, the FlashInfer baseline solution of the
contest dataset, which concatenates its inputs on every call and does not
reflect the library kernel's own speed; for MoE, FlashInfer
\texttt{trtllm\_fp8\_block\_scale\_moe}. The MoE Opus 5 whole-definition
solution is bimodal across rounds on the single-token workload (63--122
$\mu$s); the median is shown. The sessions' own reports are generally more
favorable than these numbers: the GDN Opus 4.8 session read a millisecond
timer as microseconds, the MoE Opus 4.8 session's benchmark hit a CUDA graph
cached on input pointers, and the MoE Opus 5 session timed itself on inputs
it generated. These rows await the same-GPU paired re-measurement of
Section~\ref{sec:protocol}.

\begin{table}[t]
\centering
\caption{Per-workload device time behind Table~\ref{tab:scope} ($\mu$s, lower is better). Each cell is Opus 4.8 / Opus 5 / Fable 5.1.}
\label{tab:scope_us}
\small
\setlength{\tabcolsep}{4pt}
\resizebox{\linewidth}{!}{%
\begin{tabular}{@{}llllrr@{}}
\toprule
Operator & Shape & JustGoal, whole definition & KDA-0.5 reproduction & Contest kernel & Library \\
\midrule
FA4 backward & MHA & 2675 / 1702 / 1932 & 1337 / 1416 / 1200 & --- & 1371 \\
 & GQA & 2701 / 1647 / 1565 & 1238 / 1478 / 1163 & --- & 1456 \\
 & varlen MHA & 3426 / 1985 / 2110 & 1557 / 1697 / 1378 & --- & 1572 \\
\addlinespace
DSA forward & heavy load & 19.3 / 9.58 / 13.3 & 14.1 / 8.30 / 7.43 & 10.13 & 585 \\
 & small request & 10.0 / 4.02 / 4.73 & 5.93 / 4.41 / 3.56 & 3.36 & 661 \\
 & mixed lengths & 13.5 / 6.90 / 9.30 & 9.22 / 6.29 / 5.11 & 6.92 & 652 \\
\addlinespace
GDN forward & long sequence & 1126 / 213 / 178 & 710 / 126 / 234 & 124.9 & --- \\
 & single short seq. & 13.96 / 5.96 / 5.36 & 15.2 / 4.74 / 4.67 & 4.81 & --- \\
 & few long seqs. & 1863 / 198 / 203 & 973 / 138 / 312 & 107.7 & --- \\
\addlinespace
MoE FP8 & single token & 294 / 300 / 53.7 & 135 / 72.9 / 42.1 & 52.0 & 149 \\
 & medium batch & 551 / 433 / 431 & 336 / 386 / 280 & 275.8 & 435 \\
 & large prefill & 2703 / 1390 / 2431 & 1271 / 1582 / 923 & 1028.5 & 2515 \\
\bottomrule
\end{tabular}}
\end{table}

Correctness was re-checked under the protocol of Section~\ref{sec:protocol}:
FA4 uses the official FlashAttention standard, and every other operator uses
the FlashInfer-Bench standard. The whole-definition solutions of all twelve
operator-model combinations (fifteen directories, since FA4 backward has a
fixed-length and a varlen definition) pass on every timed workload, with
three exceptions: the Opus 4.8 and Opus 5 DSA solutions exceed
the lse tolerance by 1.4--1.7$\times$ on the small-request workload, where
KDA-0.5's lse is almost exact and the tolerance is tiny, both absolute errors
being below $10^{-5}$; and the Opus 4.8 MoE solution gets 3 rows wrong as a
whole on large prefill, apparently near-ties in expert selection, with the
mean error equal to KDA-0.5's, which the official evaluator passes through its
90\% matched-ratio rule. In the KDA-0.5 reproduction, the Opus 4.8 DSA
solution returns only the attention output and no lse, which the official
evaluator did not catch; the Opus 4.8 MoE solution calls FlashInfer library
kernels and fails the standard above on two further workloads. Two items
failed the public-repository retest as released: the three Opus 4.8 DSA KDA-0.5
reproduction implementations matched an earlier failing upstream version, and
the Opus 5 MoE whole-definition solution hit a Triton launch-parameter
mismatch on all 19 workloads (36 parameters expected, 27 given).

\end{document}